\documentclass[twocolumn,showpacs,prb]{revtex4}
\usepackage{mathrsfs}
\usepackage{amsmath}
\usepackage{amssymb}
\usepackage{graphicx}
\usepackage{amsfonts}
\usepackage{txfonts}

\begin{document}

\title{Quantum many-body theory of qubit decoherence in a finite-size spin
bath}
\author{Wen Yang}
\author{Ren-Bao Liu}
\thanks{rbliu@phy.cuhk.edu.hk}
\affiliation{Department of Physics, The Chinese University of Hong Kong, Shatin, N. T.,
Hong Kong, China}
\pacs{
 76.20.+q,  03.65.Yz,  76.60.Lz,  76.30.-v        }

\begin{abstract}
Decoherence of a center spin or qubit in a spin bath is essentially
determined by the many-body bath evolution. We develop a cluster-correlation
expansion (CCE) theory for the spin bath dynamics relevant to the qubit
decoherence problem. A cluster correlation term is recursively defined as
the evolution of a group of bath spins divided by the cluster correlations
of all the subgroups. The so-defined correlation accounts for the authentic
(non-factorizable) collective excitations within a given group. The bath
propagator is the product of all possible cluster correlation terms. For a
finite-time evolution as in the qubit decoherence problem, a convergent
result can be obtained by truncating the expansion up to a certain cluster
size. The two-spin cluster truncation of the CCE corresponds to the
pair-correlation approximation developed previously [Phys. Rev. B \textbf{74}%
, 195301 (2006)]. In terms of the standard linked cluster
expansion, a cluster correlation term is the infinite summation
of all the connected diagrams with all and only the spins in the
group flip-flopped, and thus the expansion is exact whenever
converges. When the individual contribution of each higher-order
correlation term to the decoherence is small (while all the terms
combined in product could still contribute substantially), as the
usual case for relatively large baths where the decoherence could
complete well within the bath spin flip-flop time, the CCE
coincides with the cluster expansion [Phys. Rev. B \textbf{74},
035322 (2006)]. For small baths, however, the qubit decoherence
may not complete within the bath spin flip-flop timescale and
thus individual higher-order cluster correlations could grow
significant. In such cases, only the CCE converges to the exact
coherent dynamics of multi-spin clusters. We check the accuracy of
the CCE in an exactly solvable spin-chain model.
\end{abstract}

\maketitle

\section{Introduction}

The decoherence of a center spin in a spin bath\cite{Prokofev2000_RPP} has
been of interest in spin resonance spectroscopy for a long history\cite%
{Kubo1954_JPSJ,Pines1955,Feher1959,Klauder1962} and is also a paradigmatic
model in studying the state collapse in quantum mechanics. Recent revisited
interest in this problem is mostly due to the decoherence issue in quantum
computing.\cite{Loss1998,Imamoglu1999,Awschalom2002_book} Most relevant are
single electron spins in quantum dots or impurity centers where the
dominating decoherence mechanism at low temperatures (such as below a few
Kelvins) is the nuclear spins of the host
lattice.\cite{Merkulov2002,Khaetskii2002,Khaetskii2003,Coish2004,Semenov2003,
Sousa2003_PRB67,Sousa2003_PRB68,Witzel2005,Witzel2006,Witzel2007,
Witzel2007_PRBCDD, Yao2006_PRB,Yao2007_RestoreCoherence,Liu2007_NJP,
Saikin2007,Deng2006, Fujisawa2002,Elzerman2004_Nature,Kroutvar2004_Nature,
Johnson2005_Nature,Braun2005_PRL,Atature2006,Koppens2005_Science, Petta2005_Science,Greilich2006,Laird2006_PRL}
In a small system such as a quantum dot, the center spin (hereafter referred
to as qubit for clarity) and the spin bath, in the time-scale of
decoherence, is a relatively isolated subsystem in the whole environment.
Thus the qubit decoherence is due to the entanglement with the bath during
the coherent evolution of the whole system.\cite{Yao2006_PRB}

In this paper, we are interested in the so-called pure dephasing in which
the qubit experiences no longitudinal relaxation but only loses its
off-diagonal phase coherence. The pure dephasing is relevant to an electron
spin under a moderate or strong magnetic field ($\gtrsim 0.1$~T for a
typical GaAs dot), where the electron spin flip due to the hyperfine
interaction is largely suppressed by the Zeeman energy mismatch between the
electron and the nuclei. In the absence of qubit flip, a qubit-bath system
has a Hamiltonian as
\begin{equation}
H=\left\vert +\right\rangle H^{(+)}\left\langle +\right\vert +\left\vert
-\right\rangle H^{(-)}\left\langle -\right\vert ,
\label{pure_dephasing_H}
\end{equation}
by which the spin bath is driven by different Hamiltonians $H^{(\pm )}$
depending on the qubit states $\left\vert \pm \right\rangle $. When a
coherent qubit state $C_{+}\left\vert +\right\rangle +C_{-}\left\vert
-\right\rangle $ is prepared, the initial state of the qubit-bath system is
the product state $(C_{+}\left\vert +\right\rangle +C_{-}\left\vert
-\right\rangle )\otimes \left\vert \mathcal{J}\right\rangle $. At time ${T}$%
, the bath evolution $\left\vert \mathcal{J}\right\rangle \rightarrow
\left\vert \mathcal{J}^{\pm }({T})\right\rangle \equiv e^{-iH^{(\pm )}{T}%
}\left\vert \mathcal{J}\right\rangle $ predicated on the qubit state $|\pm
\rangle $ establishes an entangled state $C_{+}\left\vert +\right\rangle
\otimes \left\vert \mathcal{J}^{+}({T})\right\rangle +C_{-}\left\vert
-\right\rangle \otimes \left\vert \mathcal{J}^{-}({T})\right\rangle $. The
qubit coherence is reduced from $\rho _{+-}(0)=C_{+}C_{-}^{\ast }$ to $\rho
_{+-}({T})=C_{+}C_{-}^{\ast }\left\langle \mathcal{J}^{-}({T})|\mathcal{J}%
^{+}({T})\right\rangle $. The decoherence is characterized by the bath state
overlap $\left\langle \mathcal{J}^{-}({T})|\mathcal{J}^{+}({T})\right\rangle$.
Thus the key is the many-body bath dynamics caused by the interaction
within the bath. Without many-body interactions, the bath would not evolve
(except for a trivial phase factor that can be eliminated by standard
spin-echo\cite{Hahn1950_PR}) and the qubit coherence would not decay.

Recently, a variety of quantum many-body theories for nuclear spin bath
dynamics have been developed including the density matrix cluster expansion
(CE),\cite{Witzel2006,Witzel2007,Witzel2007_PRBCDD} the pair-correlation
approximation,\cite{Yao2006_PRB,Yao2007_RestoreCoherence,Liu2007_NJP} and
the linked-cluster expansion (LCE).\cite{Saikin2007} In the pair-correlation
approximation, each pair-wise flip-flop of nuclear spins is identified as an
elementary excitation mode and is taken as independent of each other. To
study the higher oder correlations, the Feynman diagram LCE is developed.
The evaluation of higher-order LCE, however, is rather tedious due to the
increasing number of diagrams, especially for spins higher than 1/2 (see the
Appendix for details). The density matrix CE is developed in the spirit of
the standard cluster expansion or virial expansion for interacting gases in
grand canonical ensembles.\cite{Beth1936_CE,Beth1937_CE,Kahn1938_CE} It
serves as a simple method (without the need to count or evaluate Feynman
diagrams) to include the higher-order spin interaction effects beyond the
pair-correlation approximation. The CE calculations show that when the
pair-correlation contribution to the decoherence is suppressed by pulse
control,\cite{Yao2007_RestoreCoherence} the higher-order correlations are
not negligible.\cite{Witzel2007_PRBCDD} In order to obtain the decoherence
exponent from the CE, however, the terms involving overlapping clusters have
to be neglected, which make the accuracy problematic even when the expansion
converges. This problem limits the CE to applications in large baths in
which the contribution of each individual cluster to the decoherence is
small and the overlapping correction is unimportant.

In this work, we develop a cluster-correlation expansion (CCE) method in
which the bath spin evolution are factorized into cluster correlations. Each
cluster correlation term is equivalent to the sum of all the LCE series\cite%
{Saikin2007} consisting of a given set of bath spins flip-flopped. In
particular, the two-spin cluster correlations include all diagrams with two
spins flip-flopped and is equivalent to the pair-correlation approximation.%
\cite{Yao2006_PRB,Yao2007_RestoreCoherence,Liu2007_NJP} The CCE bears the
accuracy of the LCE (the results are accurate whenever converge) and the
simplicity of the CE (without the need to count or evaluate Feynman
diagrams), while free from the large-bath restriction of the CE. The CCE  coincides with
the CE in the leading order of the short-time expansion, which is
applicable in large spin baths where the decoherence completes well within
the bath spin flip-flop time. For small baths, however, the qubit decoherence may not complete
within the bath spin flip-flop time and individual higher-order cluster
correlations could grow significant. In
this case only the CCE converges to the exact coherent dynamics of
multi-spin clusters. Such coherent dynamics of small clusters of bath spins
is of special interest in systems with randomized qubit-bath couplings. An
interesting example is nitrogen-vacancy centers in diamonds\cite%
{Jelezko2004_PRLNV,Childress2006_ScienceNV,
Dutt2007_ScienceNV,Hanson2008_ScienceNV} which are coupled to randomly
located bath spins (carbon-13 and nitrogen nuclear spins) in the
proximity.

This paper is organized as follows. In Section II, we derive, for a generic
spin bath Hamiltonian, the CCE from a recursive cluster factorization
procedure. We show that the CCE is equivalent to an infinite resummation of
the LCE series and also compare it to the CE. In section III, we check the
accuracy and convergence of the truncated CCE
in an exactly solvable model (the one-dimensional spin-1/2 XY model).
Section IV draws the conclusions. The Appendix contains the details of the
LCE for spin bath dynamics, extended to spins higher than $1/2$.

\section{Cluster-correlation expansion}

\subsection{Motivation: Pair-correlation approximation and beyond}

\label{Sec:motivation}

As discussed in the Introduction, the Hamiltonian for the pure dephasing
problem has the form of Eq.~(\ref{pure_dephasing_H}). For a given initial
bath state $\left\vert \mathcal{J}\right\rangle $ (which could be considered
as one sample chosen from a thermal ensemble), the qubit coherence is
characterized by
\begin{equation*}
{\mathcal{L}}({T})=\langle \mathcal{J}|e^{iH^{(-)}{T}}e^{-iH^{(+)}{T}}|%
\mathcal{J}\rangle .
\end{equation*}%
For a thermal ensemble of baths, a further ensemble average should be
processed. The thermal fluctuation leads to the inhomogeneous broadening,
which can be eliminated by spin echo. To focus on the qubit decoherence due
to the quantum dynamics of the bath, throughout this paper, we consider the
decoherence for a single bath state without the ensemble average. For
temperatures much higher than the bath spin flip-flop rates ($\sim $10$^{-9}$%
~K for nuclear spins in GaAs), the thermal ensemble has no off-diagonal
coherence and $\left\vert \mathcal{J}\right\rangle $ can be taken as a
noninteracting product state $\left\vert \mathcal{J}\right\rangle =\mathbf{%
\otimes }_{n}\left\vert j_{n}\right\rangle $, where $j_{n}$ denotes the
quantum number of the $n$th bath spin quantized along the external magnetic
field.

In this subsection we illustrate the central idea of the CCE method (as an
extension of the previously developed pair-correlation approximation\cite%
{Yao2006_PRB,Yao2007_RestoreCoherence,Liu2007_NJP}) using a bath consisting
of $N$ spins $\mathbf{J}_{1}$, $\mathbf{J}_{2}$, $\cdots $, and $\mathbf{J}%
_{N}$ with only pairwise secular interactions
\begin{align}
H^{(\pm )}=& \pm \frac{\Omega }{2}+\sum_{n}\left( Z_{n}\pm \frac{z_{n}}{2}%
\right) J_{n}^{z}+\sum\limits_{m\neq n}\left( D_{m,n}\pm \frac{d_{m,n}}{2}%
\right) J_{m}^{z}J_{n}^{z}  \notag \\
& +\sum\limits_{m\neq n}\left( B_{m,n}\pm \frac{b_{m,n}}{2}\right)
J_{m}^{+}J_{n}^{-},  \label{HPN}
\end{align}%
where $\Omega $ $(Z_{n})$ is the qubit (bath spin) splitting energy,
$z_{n}$ is the diagonal qubit-bath spin interaction constant
(corresponding to the hyperfine interaction strength in an
electron-nuclear spin system), $D_{n,m}$ ($B_{n,m}$) is the diagonal
(off-diagonal) intrinsic bath interaction strength, and
$d_{n,m}$ ($b_{n,m}$) is the diagonal (off-diagonal) extrinsic bath
interaction depending on the qubit states (which could result
from the interaction mediated by virtual flips of the qubit spin
while real flips are suppressed by the large energy
mismatch\cite{Yao2006_PRB}).

\begin{figure}[tbp]
\includegraphics[width=0.7\columnwidth]{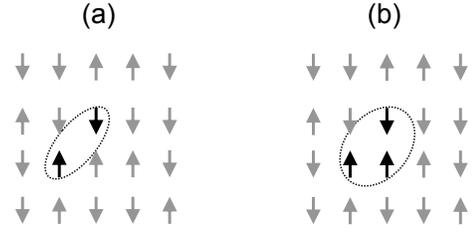}
\caption{Visualization of a cluster containing (a) two or (b) three bath
spins (black arrows). The spins outside the cluster (gray arrows) are taken
as frozen in calculating the cluster contribution.}
\label{G_Cluster23}
\end{figure}

In the pair-correlation approximation, the qubit coherence is given by the
product of all possible spin pair contributions up to a phase factor,\cite%
{Yao2006_PRB}
\begin{equation}
{\mathcal{L}}\approx \prod_{\{i,j\}}{L}_{\{i,j\}},  \label{PCA_FID}
\end{equation}%
where the contribution due to the flip-flops of a spin pair $\{i,j\}$ [see
Fig.~\ref{G_Cluster23}(a)] is
\begin{equation}
{L}_{\{i,j\}}=\left\langle \mathcal{J}\left\vert e^{ih_{\{i,j\}}^{(-)}{T}%
}e^{-ih_{\{i,j\}}^{(+)}{T}}\right\vert \mathcal{J}\right\rangle ,
\label{UIJ_FID}
\end{equation}%
with the Hamiltonian governing the pair dynamics obtained from the full
Hamiltonian $H^{(\pm )}$ by excluding the flip-flops of all spins other than
$\mathbf{J}_{i}$ and $\mathbf{J}_{j}$
\begin{align}
h_{\{i,j\}}^{(\pm )}=& \pm \frac{\Omega }{2}+\sum_{n}\left( Z_{n}\pm \frac{%
z_{n}}{2}\right) J_{n}^{z}+\sum\limits_{m\neq n}\left( D_{m,n}\pm \frac{%
d_{m,n}}{2}\right) J_{m}^{z}J_{n}^{z}  \notag \\
& +\left( B_{i,j}\pm \frac{b_{i,j}}{2}\right) \left(
J_{i}^{+}J_{j}^{-}+J_{i}^{-}J_{j}^{+}\right) ,
\end{align}%
which is equivalent to replacing in the full Hamiltonian the spins outside
the pair with their mean-field averages,
\begin{equation}
h_{\{i,j\}}^{(\pm )}\equiv H^{(\pm )}\left( \mathbf{J}_{i},\mathbf{J}%
_{j},\left\{ \left\langle \mathcal{J}\left\vert \mathbf{J}_{n\neq
i,j}\right\vert \mathcal{J}\right\rangle \right\} \right) .
\end{equation}%
The pair-correlation approximation is valid for situations where pair
correlations dominate. When the collective flip-flops of more spins become
important, we need to consider the higher-order correlation correction ${%
\mathcal{L}}_{\mathrm{corr}}$ defined by
\begin{equation}
{\mathcal{L}}=\left( \prod_{\{i,j\}}{L}_{\{i,j\}}\right) {\mathcal{L}}_{%
\mathrm{corr}}.
\end{equation}%
To illustrate how the high-order correction could be evaluated, let us
consider a spin bath of only three spins $\{1,2,3\}$. Obviously, the result
is
\begin{equation}
{\mathcal{L}}_{\mathrm{corr}}=\frac{{\mathcal{L}}}{L_{\{1,2\}}L_{\{2,3\}}L_{%
\{3,1\}}}.
\end{equation}%
This result motivates a definition of non-factorizable three-spin
correlations due to collective flip-flops. For a three-spin cluster $%
\{i,j,k\}$ in a bath [see Fig.~\ref{G_Cluster23}(b)], when all the spins
outside the cluster are frozen, the qubit coherence is
\begin{equation*}
L_{\{i,j,k\}}=\left\langle \mathcal{J}\left\vert e^{ih_{\{i,j,k\}}^{(-)}{T}%
}e^{-ih_{\{i,j,k\}}^{(+)}{T}}\right\vert \mathcal{J}\right\rangle ,
\end{equation*}%
with the cluster Hamiltonian
\begin{align*}
h_{\{i,j,k\}}^{(\pm )}\equiv & \pm \frac{\Omega }{2}+\sum_{n}\left( Z_{n}\pm
\frac{z_{n}}{2}\right) J_{n}^{z}+\sum\limits_{m\neq n}\left( D_{m,n}\pm
\frac{d_{m,n}}{2}\right) J_{m}^{z}J_{n}^{z} \\
& +\sum_{m,n\in \{i,j,k\}}\left( B_{m,n}\pm \frac{b_{m,n}}{2}\right)
J_{m}^{+}J_{n}^{-} \\
=& H^{\left( \pm \right) }\left( \mathbf{J}_{i},\mathbf{J}_{j},\mathbf{J}%
_{k},\left\{ \left\langle \mathcal{J}\left\vert \mathbf{J}_{n\notin
\{i,j,k\}}\right\vert \mathcal{J}\right\rangle \right\} \right) ,
\end{align*}%
obtained from the full Hamiltonian $H^{(\pm )}$ by replacing the spins
outside the cluster with their mean-field averages. The authentic (or
non-factorizable) three-spin correlation is singled out by excluding all the
pair-correlations
\begin{equation*}
\tilde{L}_{\{i,j,k\}}=\frac{L_{\{i,j,k\}}}{L_{\{i,j\}}L_{\{i,k\}}L_{\{j,k\}}}%
.
\end{equation*}%
If all such three-spin correlations are picked up, the qubit coherence is
given by (up to a global phase factor)
\begin{equation*}
{\mathcal{L}}\approx \prod_{\{i,j\}}L_{\{i,j\}}\cdot \prod_{\{i,j,k\}}\tilde{%
L}_{\{i,j,k\}}.
\end{equation*}%
Thus a systematic cluster correlation expansion is motivated.

\subsection{General formalism of cluster-correlation expansion}

\label{subsec_CCEDefinition}

\begin{figure}[ptb]
\includegraphics[width=\columnwidth]{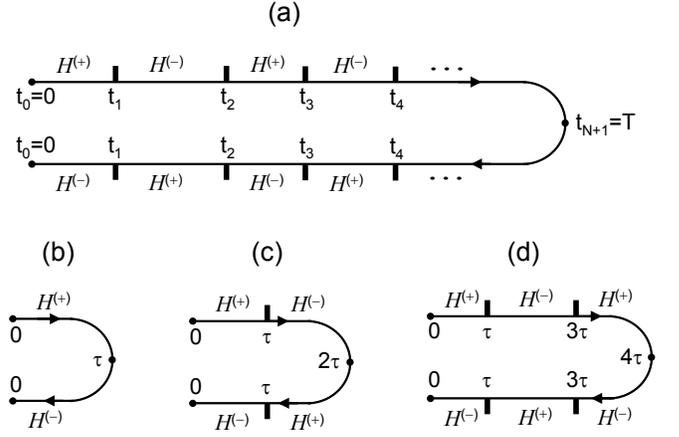}
\caption{(a) Visualization of an arbitrary sequence of controlling $\protect%
\pi$ pulses at $t_{1}$,$t_{2}$,$\cdots$, (indicated by vertical lines) and
the corresponding contour-time-dependent Hamiltonian for electron spin decoherence.
(b), (c), and (d) exemplify the cases of free-induction decay, Hahn echo,
and Carr-Purcell echo, respectively.}
\label{G_Contour}
\end{figure}

We consider a generic bath Hamiltonian for the pure dephasing problem
\begin{equation*}
H^{\left( \pm \right) }=H^{\left( \pm \right) }\left( {\mathbf{J}}_{1},{%
\mathbf{J}}_{2},\cdots ,{\mathbf{J}}_{N}\right) ,
\end{equation*}%
which need not contain only pairwise interactions, or conserve the spin
angular momentum along any direction. For instance, multi-spin interaction
terms like $J_{i}^{z}J_{j}^{+}J_{k}^{-}$ and non-secular terms like $%
J_{i}^{+}J_{j}^{z}$ could be present.

For qubit decoherence under the control of an arbitrary sequence of $\pi $%
-pulses applied at $t_{1}$, $t_{2}$, $\cdots $, as shown schematically in
Fig.~\ref{G_Contour}(a), the bath evolution predicated on the qubit state is
given by
\begin{equation*}
\left\vert \mathcal{J}^{\pm }({T})\right\rangle ={U}^{(\pm )}\left\vert
\mathcal{J}\right\rangle ,
\end{equation*}%
where
\begin{equation}
{U}^{(\pm )}\equiv \cdots e^{-iH^{(\pm )}\left( t_{3}-t_{2}\right)
}e^{-iH^{(\mp )}\left( t_{2}-t_{1}\right) }e^{-iH^{(\pm )}t_{1}}.
\label{JNP}
\end{equation}%
The qubit coherence at the end of the evolution is
\begin{equation}
{\mathcal{L}}=\left\langle \mathcal{J}\left\vert \left( {U}^{(-)}\right)
^{\dagger }{U}^{(+)}\right\vert \mathcal{J}\right\rangle .
\label{U_ARBITRARY}
\end{equation}%
It can be written in the contour time-ordered form as\cite{Saikin2007}
\begin{equation}
{\mathcal{L}}=\left\langle \mathcal{J}\left\vert \mathscr{T}_{\mathrm{c}%
}e^{-i\int_{\text{c}}H(t)dt}\right\vert \mathcal{J}\right\rangle ,
\label{W0TimeOrdered}
\end{equation}%
where $\mathscr{T}_{\mathrm{c}}$ is the time-ordering operator along the
contour $C:0\rightarrow T\rightarrow 0$. As illustrated in Fig.~\ref%
{G_Contour}(a), the contour time-dependent Hamiltonian $H(t)$ alternatively
switches between $H^{(+)}$ and $H^{(-)}$ each time the qubit state is
flipped by a $\pi $-pulse or when the time direction is reverted at $T$. The
examples for free-induction decay, single-pulse Hahn echo, and Carr-Purcell
echo are illustrated in Figs.~\ref{G_Contour} (b), (c), and (d),
respectively.

Following the idea illustrated in the previous subsection, the cluster
correlations are recursively defined as follows.

\begin{enumerate}
\item The empty-cluster correlation
\begin{equation*}
\tilde{L}_{\varnothing }\equiv L_{\varnothing }\equiv e^{-i\int_{\text{c}%
}\left\langle \mathcal{J}\left\vert {H}(t)\right\vert \mathcal{J}%
\right\rangle dt}
\end{equation*}%
is a pure phase factor obtained from Eq.~(\ref{W0TimeOrdered}) by replacing
the bath Hamiltonian $H(t)$ with its mean-field average $\left\langle
\mathcal{J}\left\vert {H}(t)\right\vert \mathcal{J}\right\rangle$.

\item The single-spin correlation
\begin{equation*}
\tilde{L}_{\{i\}}\equiv L_{\{i\}}/\tilde{L}_{\varnothing},
\end{equation*}
where
\begin{equation*}
L_{\{i\}}\equiv \left \langle \mathcal{J}\left \vert \mathscr{T}_{\mathrm{c}%
}e^{-i\int_{\text c}{h}_{\{i\}}(t)dt}\right \vert \mathcal{J}\right \rangle
\end{equation*}
is obtained from Eq.~(\ref{W0TimeOrdered}) by replacing the bath Hamiltonian
$H(\{ \mathbf{J}_{n}\},t)$ with
\begin{equation*}
h_{\{i\}}(t)\equiv H\left( \mathbf{J}_{i},\left \{ \left \langle \mathcal{J}%
\left \vert \mathbf{J}_{n\neq i}\right \vert \mathcal{J}\right \rangle
\right \} ,t\right) .
\end{equation*}
in which all the spin operators except ${\mathbf{J}}_{i}$ are mean-field
averaged.

\item The two-spin (pair) correlation
\begin{equation*}
\tilde{L}_{\{i,j\}}\equiv L_{\{i,j\}}/\left( \tilde{L}_{\varnothing }\tilde{L%
}_{\{i\}}\tilde{L}_{\{j\}}\right) ,
\end{equation*}%
where
\begin{equation*}
L_{\{i,j\}}\equiv \left\langle \mathcal{J}\left\vert \mathscr{T}_{\mathrm{c}%
}e^{-i\int_{\text{c}}{h}_{\{i,j\}}(t)dt}\right\vert \mathcal{J}\right\rangle
\end{equation*}%
is obtained from Eq.~(\ref{W0TimeOrdered}) by replacing the bath Hamiltonian
$H(\{\mathbf{J}_{n}\},t)$ with
\begin{equation*}
h_{\{i,j\}}(t)\equiv H\left( \mathbf{J}_{i},\mathbf{J}_{j},\left\{
\left\langle \mathcal{J}\left\vert \mathbf{J}_{n\notin \{i,j\}}\right\vert
\mathcal{J}\right\rangle \right\} ,t\right),
\end{equation*}
in which all the spin operators except ${\mathbf{J}}_{i}$ and ${\mathbf{J}}%
_{j}$ are mean-field averaged.

\item So on and so forth, the cluster correlation for an arbitrary set of
bath spins $\mathcal{C}$ is defined as
\begin{equation}
\tilde{L}_{\mathcal{C}}\equiv \frac{L_{\mathcal{C}}}{\prod \limits_{\mathcal{%
C}^{\prime}\subset \mathcal{C}}\tilde{L}_{\mathcal{C}^{\prime}}},
\label{UC_EXPANSION}
\end{equation}
where
\begin{equation}
L_{\mathcal{C}}\equiv \left \langle {\mathcal{J}}\left \vert \mathscr{T}_{%
\mathrm{c}}e^{-i\int_{\text c}{h}_{\mathcal{C}}{}(t)dt}\right \vert {%
\mathcal{J}}\right \rangle  \label{UC_DEF}
\end{equation}
is obtained from Eq.~(\ref{W0TimeOrdered}) by replacing the bath Hamiltonian
$H(\{ \mathbf{J}_{n}\},t)$ with
\begin{equation}
h_{\mathcal{C}}(t)\equiv H\left( \left \{ \mathbf{J}_{n\in \mathcal{C}%
}\right \} ,\left \{ \left \langle \mathcal{J}\left \vert \mathbf{J}%
_{n\notin \mathcal{C}}\right \vert \mathcal{J}\right \rangle \right \}
,t\right) ,  \label{H_cluster}
\end{equation}
in which all the spin operators outside the cluster are mean-field averaged
or their flip-flops are frozen.
\end{enumerate}

Thus, by definition, the qubit coherence is factorized into all possible
cluster correlations as
\begin{equation}
{\mathcal{L}}=\prod_{\mathcal{C}\subseteq \{1,2,\cdots ,N\}}\tilde{L}_{%
\mathcal{C}}.  \label{CCE_EXACT}
\end{equation}%
Calculating the CCE to the maximum order $\tilde{L}_{\{1,2,\cdots ,N\}}$
amounts to solving the exact bath propagator, which is in general not
possible. In the decoherence problem, we consider a finite-time evolution
and it often suffices to truncate the expansion by keeping cluster
correlations up to a certain size $M$, as the $M$th-order truncation of the
CCE ($M$-CCE for short),
\begin{equation}
{\mathcal{L}}^{\left( M\right) }=\prod_{\left\vert \mathcal{C}\right\vert
\leq M}\tilde{L}_{\mathcal{C}},  \label{CCE_MTH}
\end{equation}%
where $\left\vert \mathcal{C}\right\vert $ is the number of spins contained
in the cluster $\mathcal{C}$. As an example, for the pairwise Hamiltonian in
Eq.~(\ref{HPN}), the lowest nontrivial order of truncation is
\begin{equation}
{\mathcal{L}}^{\left( 2\right) }=\tilde{L}_{\varnothing }\prod_{\{i,j\}}%
\tilde{L}_{\{i,j\}}=\tilde{L}_{\varnothing }\prod_{\{i,j\}}\left(
L_{\{i,j\}}/\tilde{L}_{\varnothing }\right) ,  \label{CCE_PCA}
\end{equation}%
which is the pair-correlation approximation.

\subsection{Relation to linked-cluster expansion}

\label{subsec_LCE}

Saikin \textit{et al.} have recently developed an LCE method for the qubit
decoherence in a spin-1/2 bath.\cite{Saikin2007} The detailed descriptions
of the LCE for a generic spin bath are given in the Appendix. In general,
the bath evolution can be factorized using Feynman diagrams so that
\begin{equation}
{\mathcal{L}}=\exp \left( \pi \right) ,
\end{equation}%
where
\begin{equation}
\pi =\left\langle \mathcal{J}\left\vert \mathscr{T}_{\mathrm{c}}e^{-i\int_{%
\text{c}}{H{}}(t)dt}\right\vert \mathcal{J}\right\rangle _{\mathrm{connected}%
},  \label{LCE_VK}
\end{equation}%
is the sum of all connected Feynman diagrams obtained by using Wick's
theorem on the series expansion of Eq.~(\ref{W0TimeOrdered}). Some of the
connected diagrams up to the 4th-order have been evaluated in Ref.~[%
\onlinecite{Saikin2007}] for free-induction decay and single-pulse Hahn echo
with a spin-1/2 bath Hamiltonian in the form of Eq.~(\ref{HPN}). The
complexity for diagram counting and evaluation increases dramatically when
considering higher-order diagrams (see Fig.~\ref{G_D123} in Appendix~\ref%
{append_LCEspin12}) or higher spins [see Fig.~\ref{G_Spin1LD12}(b) in
Appendix~\ref{append_LCEgeneral} for the second-order diagrams for a spin-1
bath].

\begin{figure}[tbp]
\includegraphics[width=\columnwidth]{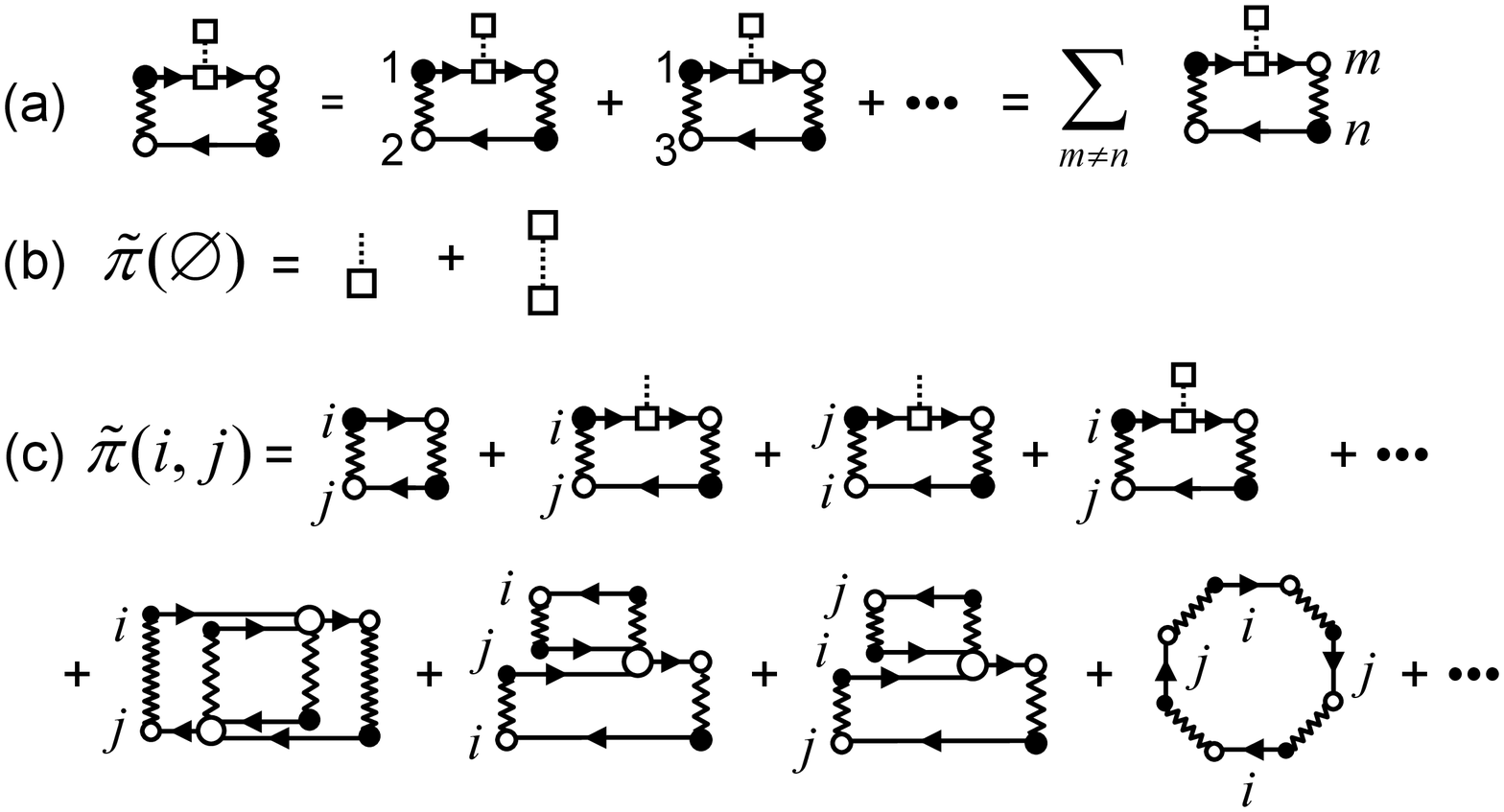}
\caption{(a) Expansion of a third-order connected diagram into diagrams
involving the flip-flops of different clusters of spins. (b) and (c) show
the diagrams contained in $\tilde{\protect\pi}(\varnothing )$ and $\tilde{%
\protect\pi}(i,j)$, respectively. In the diagrams, a solid arrow
denotes the propagation of a spin. A wavy (dotted) line connecting
two solid arrows/spins denotes a pairwise off-diagonal (diagonal)
interaction, and an open-ended line represents the interaction
with the qubit spin or an external field. Each filled circle/empty
circle/empty square on a solid arrow denotes a $J^{+}/J^{-}/J^{z}$
operator, which raises/lowers/keeps invariant the $J^{z}$ quantum
number of that spin.} \label{G_CCEU02}
\end{figure}

We notice that each diagram can be expanded as the sum of diagrams involving
the flip-flops of different clusters of spins. As an example shown in Fig.~%
\ref{G_CCEU02}(a), a third-order diagram involving the flip-flop of a spin
pair contains diagrams for spin clusters $(1,2)$, $(1,3)$, $\cdots $, where
the numbers stand for the indices of the spins flip-flopped (i.e., the $%
J^{z} $ quantum number changed). Thus all the connected diagrams can be
classified according to the spin clusters instead of the interaction orders.
For an arbitrary cluster $\mathcal{C}$, we define $\tilde{\pi}({\mathcal{C}}%
) $ as the sum of all connected diagrams in which all and only the spins in
cluster $\mathcal{C}$ have been flip-flopped. For instance, some of the
diagrams constituting $\tilde{\pi}(\varnothing )$ and $\tilde{\pi}(i,j)$ for
a spin-1/2 Hamiltonian in Eq.~(\ref{HPN}) are shown in Figs.~\ref{G_CCEU02}%
(b) and (c), respectively. With these $\{\tilde{\pi}\}$ functions, the LCE
is expressed as
\begin{equation}
\pi =\sum\limits_{\mathcal{C}\subseteq \{1,2,\cdots ,N\}}\tilde{\pi}\left( {%
\mathcal{C}}\right) .  \label{LCE_EXPANSION}
\end{equation}%
In particular, the infinite summation of all the connected diagrams for a
certain cluster ${\mathcal{C}}$ and all its subsets
\begin{equation}
\pi \left( {\mathcal{C}}\right) \equiv \sum_{{\mathcal{C}}^{\prime}
\subseteq {\mathcal{C}}}\tilde{\pi}\left( {\mathcal{C}}^{\prime }\right) ,
\end{equation}%
can be obtained from the series expansion Eq.~(\ref{LCE_VK}) by dropping all
the terms involving the flip-flop of spins outside the cluster ${\mathcal{C}}
$, or, equivalently, by reducing the bath Hamiltonian $H(t)$ to the cluster
Hamiltonian $h_{\mathcal{C}}(t)$ in which the spins outside the cluster are
mean-field averaged. Thus we have
\begin{equation}
e^{\pi \left( {\mathcal{C}}\right) }=\prod_{{\mathcal{C}}^{\prime }\subseteq
{\mathcal{C}}}e^{\tilde{\pi}\left( {\mathcal{C}^{\prime }}\right)
}=\left\langle \mathcal{J}\left\vert {\mathscr{T}}_{\text{c}}e^{-i\int_{%
\text{c}}h_{\mathcal{C}}(t)dt}\right\vert \mathcal{J}\right\rangle =L_{%
\mathcal{C}}.
\end{equation}%
Comparing this to Eqs.~(\ref{UC_EXPANSION})-(\ref{H_cluster}), we
immediately have
\begin{equation}
\tilde{L}_{\mathcal{C}}=e^{\tilde{\pi}\left( {\mathcal{C}}\right) },
\label{CCE_LCE}
\end{equation}%
i.e., a cluster correlation term corresponds to the infinite partial summation of
all the connected diagrams in which all and only the spins in the cluster
have been flip-flopped.

From the LCE expression of the CCE in Eq.~(\ref{CCE_LCE}), it is obvious
that the short-time profile of the decoherence due to clusters of a certain
size is determined by the lowest-order diagram. In particular, for
free-induction decay with the secular pair-interaction Hamiltonian in Eq.~(%
\ref{HPN}), the leading order contribution from a cluster of size $M$ is
\begin{equation}
\tilde{\pi}\left( {\mathcal{C}}\right) \sim B^{M}T^{M},
\end{equation}%
where $B$ is the typical magnitude of the pair flip-flop interaction
strength $B_{m,n}\pm b_{m,n}$. If each spin interacts, on average, with $q$
spins, then the number of size-$M$ clusters is $\sim Nq^{M-1},$ with $N$ the
total number of bath spins. The sum of all the leading order $M$-spin
connected diagrams is $\sim q^{-1}N\cdot (qBT)^{M}$. Thus for $qBT\ll 1,$
i.e., for a time $T$ much shorter than the bath spin flip-flop timescale $%
1/B $, a truncated CCE converges. The short time condition $T\ll B^{-1}$ is
usually satisfied for electron spin decoherence caused by nuclear spins in
typical quantum dots. The convergence of the truncated CCE, however, could
go well beyond the short time restriction. One such scenario is small spin
baths with disorder in qubit-bath couplings or in spin splitting energies,
in which multi-spin correlation could develop at a time well beyond the short-time
limit but the size of the contributing clusters could remain bounded due to
the localization effect in a disordered system, as will be verified later in
this paper by numerical simulations.

\subsection{Relation to cluster expansion}

\label{CCE_CE}

Witzel \textit{et al.} recently developed a density matrix CE approach to
solving the nuclear spin dynamics in the electron spin decoherence problem,%
\cite{Witzel2006,Witzel2007_PRBCDD} in the spirit of the cluster or virial
expansion for interacting gases in grand canonical ensembles.\cite%
{Beth1936_CE,Beth1937_CE,Kahn1938_CE} Below we reproduce the basic procedure
of the CE and compare it to the CCE. Instead of the ensemble CE in Ref.~[%
\onlinecite{Witzel2006}], we consider a single sample state of the bath for
a direct comparison. Defining the qubit decoherence due to a cluster ${%
\mathcal{C}}$ of bath spins as
\begin{equation}
W(\mathcal{C})\equiv \left\vert \left\langle \mathcal{J}\left\vert %
\mathscr{T}_{\mathrm{c}}e^{-i\int_{\text{c}}{h{}}_{\mathcal{C}%
}(t)dt}\right\vert \mathcal{J}\right\rangle \right\vert =\left\vert L_{%
\mathcal{C}}\right\vert ,
\end{equation}%
a hierarchy of cluster terms $\{\tilde{W}({\mathcal{C}})\}$ are recursively
defined as
\begin{subequations}
\label{CE_summation}
\begin{align}
W(i)=& \tilde{W}(i), \\
W(i,j)=& \tilde{W}(i,j)+\tilde{W}(i)\tilde{W}(j), \\
W(i,j,k)=& \tilde{W}(i,j,k)+\tilde{W}(i)\tilde{W}(j)\tilde{W}(k)+\tilde{W}%
(i,j)\tilde{W}(k)  \notag \\
& +\tilde{W}(i,k)\tilde{W}(j)+\tilde{W}(i)\tilde{W}(j,k), \\
\cdots & \cdots  \notag \\
W(\mathcal{C})=& \tilde{W}(\mathcal{C})+\sum_{\{\mathcal{C}_{j}\}}\prod_{%
\mathcal{C}_{j}}\tilde{W}(\mathcal{C}_{j}),
\end{align}%
where in the last line the sum runs over all possible partitions of the
cluster $\mathcal{C}$ into nonoverlapping nonempty subsets $\mathcal{C}_{1},%
\mathcal{C}_{2},\cdots $. The $M$th-order truncated CE ($M$-CE for short) is
\end{subequations}
\begin{equation}
W^{(M)}=\sum_{\{\mathcal{C}_{j}\},\left\vert {\mathcal{C}}_{j}\right\vert
\leq M}\prod_{\mathcal{C}_{j}}\tilde{W}(\mathcal{C}_{j}),  \label{CE_Mth}
\end{equation}%
with the sum running over all possible partitions of the bath into
nonoverlapping nonempty clusters $\mathcal{C}_{1},\mathcal{C}_{2},\cdots $
of size up to $M$.

In the cluster expansion for interacting gases in grand canonical
ensembles with translational symmetry, the contribution from
different clusters can be
factorized\cite{Beth1936_CE,Beth1937_CE,Kahn1938_CE} and the
evaluation of a truncated CE amounts to the calculation of a finite
number of finite-size cluster contributions, similar to the CCE in
this paper. For a finite-size spin bath or for a bath without
translational symmetry, however, such factorization of different
clusters does not exist, which makes it essentially impossible to
calculate the sum in Eq.~(\ref{CE_Mth}) even for a small-$M$-CE. For
example, for a bath of $N$ spins, the number of all terms containing
only pair clusters is $O(N!!)$ and all such terms have to be
individually calculated and summed in the 2-CE, which is practically
impossible.

A remedy is possible when all the cluster terms $\tilde{W}({\mathcal{C}})$
are individually small (but the sum could still be substantial). Under such
a condition, the CE can be approximated by a factorized form by adding some
overlapping terms which are higher-order small.\cite{Witzel2006} For
example, with the secular pair-interaction Hamiltonian in Eq.~(\ref{HPN}),
for which $W(i)=\tilde{W}(i)=1$, the $M$-CE is
\begin{equation}
W^{(M)}\approx \prod_{1<\left\vert \mathcal{C}\right\vert \leq M}\left[ 1+%
\tilde{W}(\mathcal{C})\right] ,  \label{CE_factorized}
\end{equation}%
under the small-term condition%
\begin{equation}
\left\vert \tilde{W}({\mathcal{C}})\right\vert \ll 1,\ \ \mathrm{for}\ \ |{%
\mathcal{C}}|>1.  \label{smallterm}
\end{equation}%
Comparing the factorization approximation in Eq.~(\ref{CE_factorized}) to
the exact $M$-CE in Eq.~(\ref{CE_Mth}), the error added is
\begin{equation}
\delta {W}^{(M)}=\sum_{i<j<k}\tilde{W}(i,j)\tilde{W}(j,k)+\sum_{i<j<k<l}%
\tilde{W}(i,j,k)\tilde{W}(k,l)+\cdots ,  \label{CE_Overlap}
\end{equation}%
containing products of any set of cluster terms sharing at least one spin,
i.e., the overlapping terms. Such overlapping terms are higher-order small
if each individual cluster term (for $|{\mathcal{C}}|>1$) is small.
Furthermore, under the small-term condition, $1+\tilde{W}({\mathcal{C}}%
)\approx \exp \left[ \tilde{W}({\mathcal{C}})\right] $, and the CE for an
arbitrary cluster becomes
\begin{equation}
W(\mathcal{C})\approx \prod_{\mathcal{C}^{\prime }\subseteq \mathcal{C}}e^{%
\tilde{W}(\mathcal{C}^{\prime })},
\end{equation}%
which takes the same form as the CCE.

Thus the CCE coincides with the CE under the small-term condition Eq.~(\ref%
{smallterm}), which is justified for large spin baths where the number of
contributing clusters is large and hence the contribution from each
individual cluster remains small within the timescale of decoherence. The
problem with the neglected overlapping terms is relevant for small spin
baths where the coherent dynamics of a small number of multi-spin clusters
dominating the decoherence may persist well beyond the bath spin flip-flop
time and the small-term condition is no longer satisfied. In this case the CE will not
converge to the exact multi-spin cluster dynamics, as will be seen in
the numerical check in the next section.

\section{Numerical check}

\label{Sec:numerical}

Here we consider an exactly solvable spin bath model (the one-dimensional
spin-1/2 XY model) and compare the exact solution to the results obtained
with the CCE and the CE. The $N$-spin bath Hamiltonian conditioned on the
qubit state $|\pm \rangle $ (with spin splitting constants dropped) is
\begin{equation}
H^{(\pm )}=\pm \sum_{n=1}^{N}\frac{z_{n}}{2}J_{n}^{z}+\sum_{n=1}^{N-1}\left(
B_{n}\pm \frac{b_{n}}{2}\right) \left(
J_{n+1}^{+}J_{n}^{-}+J_{n}^{+}J_{n+1}^{-}\right) ,  \label{HNPxymodel}
\end{equation}%
where $z_{n}$ denotes the qubit-bath spin interaction coefficient, $B_{n}$
is the intrinsic bath interaction strength, and $b_{n}$ is the
interaction dependent on the qubit state. The initial bath state $\left\vert
\mathcal{J}\right\rangle $ is taken as a product state of all bath spins, in
which the orientation of each bath spin is randomly chosen as up or down.
The qubit-bath interaction coefficients $\{z_{n}\}$ are either taken from a
sinusoidal distribution $z_{n}=z_{\mathrm{max}}\sin (n\pi /N)$ (referred to
as a ``sinusoidal'' spin chain) or randomly chosen from $\left[ 0,z_{\mathrm{max}}%
\right] $ (referred to as a ``random'' spin chain). Hereafter
$z_{\mathrm{max}}$ is taken as the unit of energy. The spin-flip
interaction strengths $\{B_{n}\}$ and $\{b_{n}\}$ are randomly
chosen from $[10^{-3},2\times 10^{-3}]$, corresponding to typical
bath spin flip-flop time $\tau _{\mathrm{sf}}\sim 10^3$. The exact
solution is obtained by the Jordan-Wigner transformation, which
transforms the interacting spin-1/2 chain to a noninteracting
fermion system.\cite{Lieb1961,Huang2006}

\subsection{Large spin bath}

\begin{figure}[tbp]
\includegraphics[width=\columnwidth]{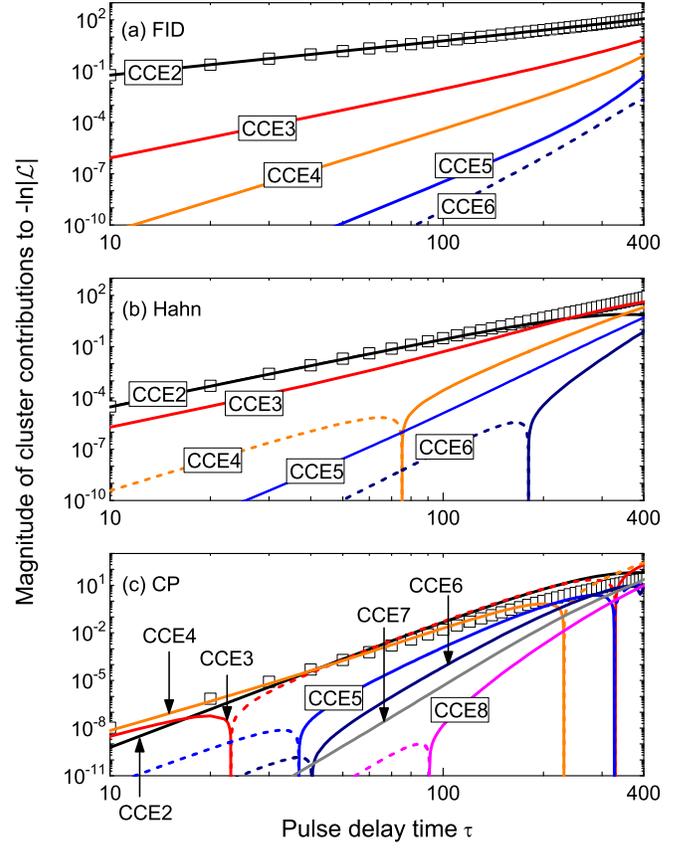}
\caption{(Color online) The magnitude of size-$m$ cluster contributions
(denoted as CCE$m$) to the exponential decoherence factor $-\ln \left\vert {%
\mathcal{L}}\right\vert $ in (a) free-induction decay (FID), (b)
Hahn echo, and (c) Carr-Purcell echo (CP) for a long ``sinusoidal''
spin chain with $N=1000$
spins. A solid (dashed) line indicates that the cluster contribution to $%
-\ln \left\vert {\mathcal{L}}\right\vert $ is positive (negative). The exact
$-\ln \left\vert \mathcal{L}\right\vert $ is also shown (empty squares) for
comparison.}
\label{G_Sine1000log}
\end{figure}

As the first example, we consider a sinusoisal chain of $N=1000$ spins. In
this case, the energy cost $\left\vert z_{n}-z_{n+1}\right\vert $ for a
pairwise flip-flop between neighboring spins varies smoothly from one end to
the other of the chain, so that a correlated cluster can grow to larger and
larger size as time passes by. For a time greater than the bath spin
flip-flop time $\tau _{\mathrm{sf}}$, the whole bath could become
correlated. But the qubit decoherence would be completed within a time $T$
much shorter than $\tau _{\mathrm{sf}}$, if the
bath size is relatively large, or
\begin{equation}
NT^{2}\tau _{\mathrm{sf}}^{-2}\gg 1,
\end{equation}%
due to the large number of small clusters contributing to the
decoherence. Under the short-time condition, the CCE can be
truncated with a rather small cut-off size $M$. The total
contributions from clusters of various sizes to the decay of the
qubit coherence as functions of the pulse delay time $\tau $ [see
Figs.~\ref{G_Contour} (b)-(d)] are shown in
Fig.~\ref{G_Sine1000log}. For the free-induction decay in
Fig.~\ref{G_Sine1000log}(a), the cluster contributions decrease
rapidly with increasing the cluster size $m$ at a time much shorter
than $\tau _{\mathrm{sf}}$, so the 2-CCE (pair correlation
approximation) already converges to the exact result. In the single-pulse
Hahn echo [Fig.~\ref{G_Sine1000log}(b)], the higher-order
correlations are more noticeable than in the free-induction decay,
but the pair correlations still dominate for $\tau \ll \tau
_{\mathrm{sf}}$. For the two-pulse Carr-Purcell echo
[Fig.~\ref{G_Sine1000log}(c)], as the decoherence due to the pair
correlations is eliminated in the leading order of the spin-flip
interactions,\cite{Witzel2007_PRBCDD} the larger-size cluster
correlations become important and a 6-CCE is required to reproduce
the exact solution.

\begin{figure}[tbp]
\includegraphics[width=\columnwidth]{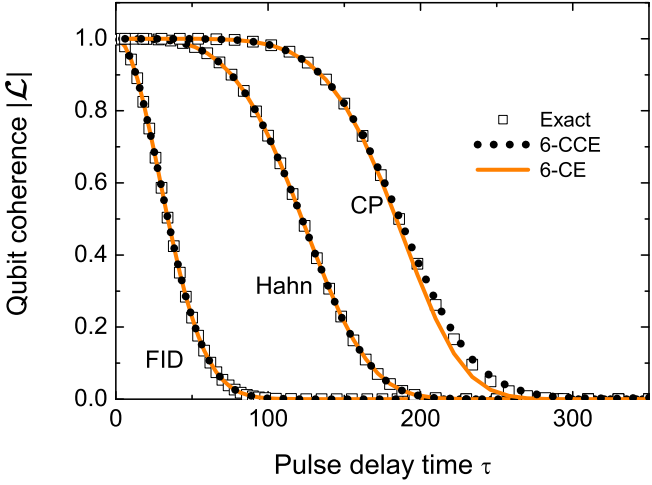}
\caption{ (Color online). Qubit coherence (empty squares) in
free-induction decay (FID),  Hahn echo, and Carr-Purcell echo (CP)
for a long ``sinusoidal'' spin chain with $N=1000$ spins. The
results from the 6-CCE (dotted lines) and 6-CE (solid lines) are
compared with the exact solutions (empty squares).}
\label{G_FIDHAHNCP}
\end{figure}

In the relatively large bath, due to the large number of contributing
clusters, the qubit decoherence completes when each individual cluster
contribution is small, i.e., the small-term condition in Eq.~(\ref{smallterm})
is satisfied. Thus the CCE coincides with the CE. This is verified in
Fig.~\ref{G_FIDHAHNCP}, where the exact solution for qubit decoherence agrees
with both the CCE and the CE truncated at the 6th order.

\subsection{Small spin bath}

\begin{figure}[tbp]
\includegraphics[width=\columnwidth]{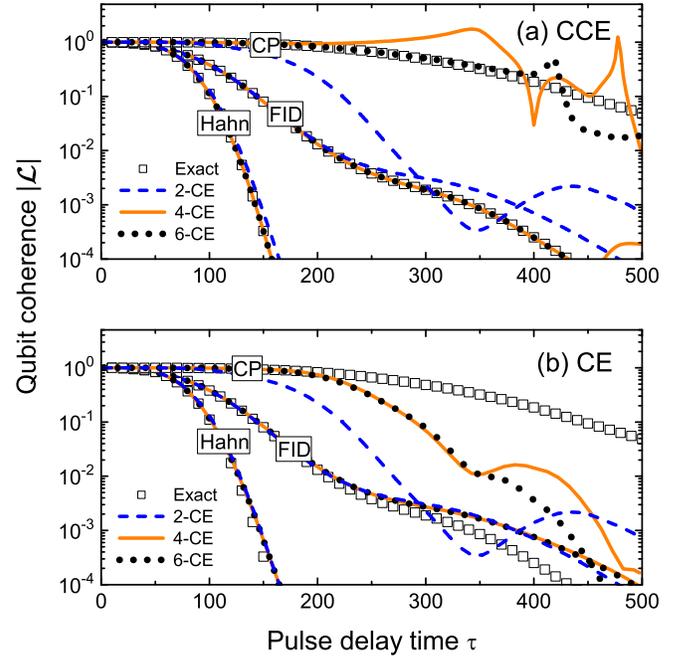}
\caption{(Color online) Qubit coherence for a short ``sinusoidal''
spin chain of $N=100$ spins in free-induction decay (FID), Hahn
echo, and Carr-Purcell echo (CP). The results from (a) CCE and (b)
CE truncated to the 2nd (dashed lines), the 4th (solid lines) and
the 6th (dotted lines) order are compared to the exact solution
(empty squares).} \label{G_CompareSine100}
\end{figure}
As the second example, we consider a short ``sinusoidal'' spin chain
consisting of $N=100$ spins. In this case the number of contributing
clusters is small and the decoherence proceeds much slower.
Fig.~\ref{G_CompareSine100} shows that both the CCE and the CE
converge to the exact result for a time much shorter than $\tau
_{{\mathrm{sf}}}$. As the time approaches and goes beyond the bath
spin flip-flop time $\tau _{{\mathrm{sf}}}$, the deviation from the
exact solution is noticeable in both the CCE and the CE, indicating
the emergence of correlations for clusters larger than the
truncation cut-off size. But the CCE agrees with the exact result as
long as it converges, while the CE converges to a different result
(for $T=\tau >250$ in free-induction decay, $T=2\tau >300$ in Hahn
echo, and $T=4\tau >800$ in Carr-Purcell echo). The deviation of the
converged CE from the exact result is due to the overlapping
correction [see Eq.~(\ref{CE_Overlap})] neglected in the CE: For
$T\gtrsim \tau _{\mathrm{sf}}$, the contribution of an
individual cluster could be sizable and the small-term condition in Eq.~(\ref%
{smallterm}) is violated, so the overlapping correction has to be taken
into account.

In the example discussed above, the CCE may not converge at a long time $%
T\gtrsim \tau _{\text{s}\mathrm{f}}$. This is because the qubit-bath spin
coupling assumes a smooth sinusoidal distribution and hence the energy cost
of each neighboring pair is small and slow-varying as a function of the
pair's position along the chain. The small and slow-varying energy cost of
a pair-flip makes it possible for one pair-flip to affect its neighboring
pairs, then the next neighbors, and so on and so forth. This way large-size
cluster correlations may grow rapidly after the time goes beyond the
pair-flip time $\tau _{\mathrm{sf}}$. In a relatively small bath of smooth
qubit-bath coupling distribution, the CCE may fail to converge
in the long time limit which is of interest in
elongating the qubit coherence time by pulse control.

\begin{figure}[tbp]
\includegraphics[width=\columnwidth]{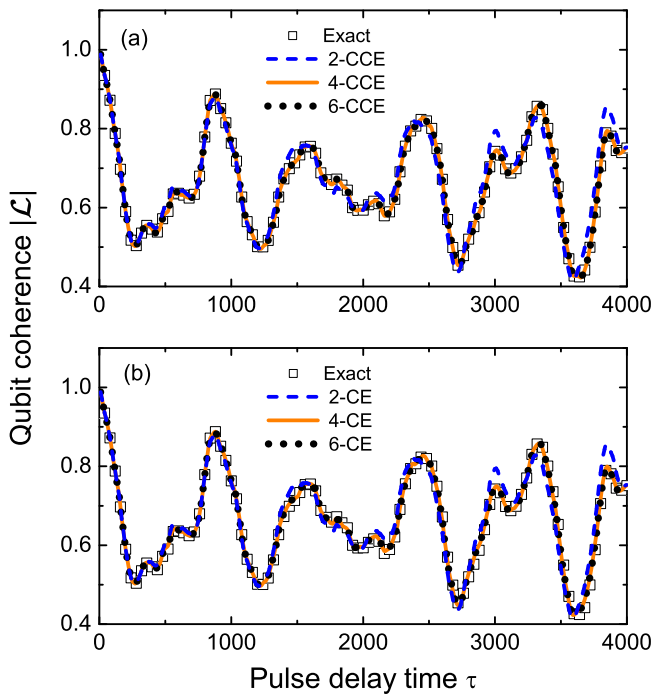}
\caption{(Color online) Qubit coherence in free induction decay for a short
``random'' spin chain with $N=100$ spins. (a) $M$-CCE and (b) $M$-CE are
compared to the exact solution (empty squares).}
\label{G_CompareRan1}
\end{figure}

\begin{figure}[tbp]
\includegraphics[width=\columnwidth]{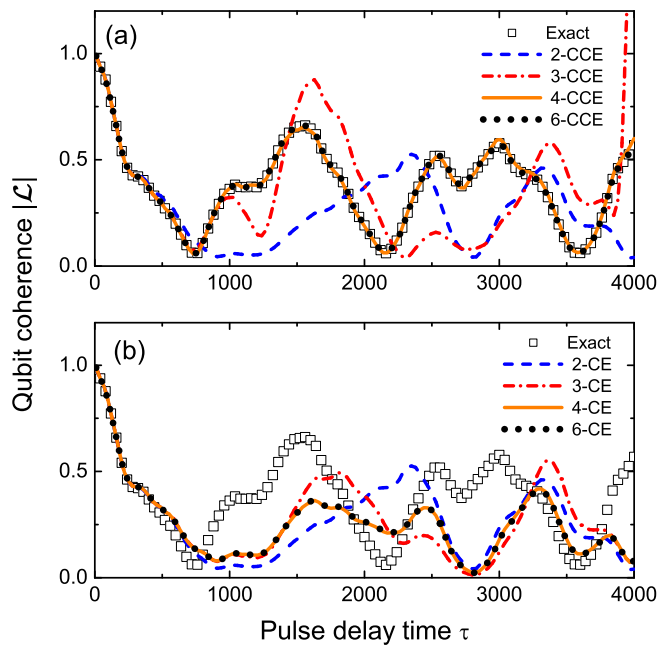}
\caption{(Color online) The same as Fig.~\protect\ref{G_CompareRan1} except
that the qubit-bath coupling constants $\{z_{n}\}$ are such that there are
four neighboring spins in near-resonance. Notice that the 4th-order and the
6th-order truncations are not distinguishable by eyes in both CCE and CE.}
\label{G_CompareRan2}
\end{figure}

The convergence problem of the CCE for small spin baths may be avoided if the qubit-bath
coupling is random so that the bath correlation is localized and the size of
correlated clusters is upper bounded. As the last example, we consider a
short ``random'' spin chain consisting of $N=100$ spins. In this case, the
pair-flip energy cost is usually much larger than the spin-flip strength
unless two neighboring spins are accidentally in near-resonance. Thus
large-size cluster correlation can hardly grow significant even for a time
well beyond the bath spin flip-flop timescale $\tau _{\mathrm{sf}}$. Figure~%
\ref{G_CompareRan1} shows that the pair-correlation approximation already suffices
for the qubit decoherence for an arbitrarily long time. Since
higher-order correlations are small, the overlapping correction to the CE
(which corresponds to at least three-spin cluster correlations) is
unimportant, good agreement between the CE and the exact solution is also
seen. Interestingly, in the small bath of random qubit-bath coupling, the
qubit decoherence is not fully developed but coherent oscillations persist
over a long time. These coherent qubit oscillations have indeed been observed
in a system of a similar nature, namely, nitrogen-vacancy centers in diamonds
coupled to nuclear spins in the proximity.\cite{Childress2006_ScienceNV}
We would like to point out that the coherent oscillations cannot be reproduced
by the LCE truncated as any finite interaction order.

As discussed in Sec.~\ref{CCE_CE}, the CCE and the CE differ in
describing the multi-spin correlations, which may be important for a
small spin bath. Such difference can indeed be seen in
Fig.~\ref{G_CompareSine100} for a short ``sinusoidal'' spin chain.
It is interesting to see the difference between the two theories in
studying the coherent multi-spin dynamics in a small bath. For this
purpose, we consider a short ($N=100$) ``random'' spin chain with
accidental near-resonance between four neighboring spins. The
pair-flip between the four spins in near-resonance may develop up to
four-spin cluster correlations as the evolution time goes beyond the
pair-flip timescale $\tau _{\mathrm{sf}}$. Indeed,
Fig.~\ref{G_CompareRan2}(a) shows coherent oscillations that are
correctly reproduced by the CCE at the 4th or higher-order
truncation. The correction by the 5th and higher-order clusters is
actually negligible, verifying that the dominating cause of the
coherent qubit oscillation is due to up to four-spin cluster
dynamics in the bath. In contrast, the CE, though already converges
at the 4th-order truncation, does not reproduce the exact solution
[see Fig.~\ref{G_CompareRan2}(b)], due to the neglect of the
overlapping terms [see Eq.~(\ref{CE_Overlap})].

\section{Conclusion}

We have developed a CCE approach to solving the many-body dynamics of a
generic interacting spin bath relevant to the center spin decoherence
problem. In this approach, the bath propagator is factorized exactly into
the product of cluster correlation terms, each of which accounts for the
correlated flip-flops of a group of bath spins. In terms of the standard
LCE, a cluster correlation term corresponds to the infinite summation of all
the connected diagrams with all and only the spins in the cluster
flip-flopped. For a finite-time evolution as in qubit decoherence, a
convergent result can be obtained by truncating the expansion up to a
certain cluster size. The CCE gives exact results whenever it converges.
The lowest nontrivial order of the CCE corresponds to the previously developed
pair-correlation approximation. Compared to the CE method, the two theories
yield similar results for large spin baths, but for small spin baths only
the CCE accurately takes into account the multi-spin cluster correlations.
As a simple method to sum over an infinite series of LCE diagrams, the CCE
method can be readily applied to the case of interacting bosons or fermions.

\acknowledgments
This work was supported by Hong Kong RGC Project 2160285.

\appendix

\section{Linked cluster expansion for spin bath dynamics}

\label{Append_LCE}

\subsection{Thermal ensemble LCE}

\label{append_LCEGaussian}

The LCE for the propagator of an arbitrary spin bath in a thermal ensemble
has been derived in Ref.~[\onlinecite{Vaks1968_LCE}]. Here we summarize the
main results including the Feynman diagram representation, as
the basis of the extension to the LCE for an arbitrary noninteracting bath
state.

For a thermal ensemble of spin baths characterized by the noninteracting
density matrix $\rho _{0}\equiv e^{-\beta H_{0}}/\mathrm{Tr}e^{-\beta H_{0}}$
with $H_{0}=\sum_{m}\omega _{m}J_{m}^{z}$, the evolution in the contour
time-ordered form is
\begin{equation}
\mathcal{L}^{\mathrm{ens}}=\mathrm{Tr}\left[ \rho _{0}\mathscr{T}_{\mathrm{c}%
}e^{-i\int_{\text{c}}{H{}}(t)dt}\right] .  \label{Propagator_ensemble}
\end{equation}%
Here the contour time-dependent Hamiltonian $H(t)$ switches alternatively
between $H^{(+)}$ and $H^{(-)}$ on the contour [see Fig.~\ref{G_Contour} (a)].
As an example, the pairwise bath Hamiltonian in
Eq.~(\ref{HPN}) leads to
\begin{align}
{H{}}(t)& =\frac{\Omega (t)}{2}+\sum\limits_{n}Z_{n}(t)J_{n}^{z}(t)+\sum%
\limits_{m\neq n}D_{m,n}(t)J_{m}^{z}(t)J_{n}^{z}(t)  \notag \\
& +\sum\limits_{m\neq n}B_{m,n}(t)J_{m}^{+}(t)J_{n}^{-}(t),
\label{HtExample}
\end{align}%
where $\Omega (t)$, $Z_{n}(t)$, $D_{m,n}(t)$, and $B_{m,n}(t)$ depend on
which contour time segment $t$ is in, and the spin operators $%
J_{n}^{z}(t)\equiv J_{n}^{z}$, $J_{n}^{\pm }(t)\equiv J_{n}^{\pm }$ are
time-independent but are written with explicit time-dependence to keep track
of the time-ordering.

Eq.~(\ref{Propagator_ensemble}) can be expanded into series as
\begin{equation}
\mathcal{L}^{\mathrm{ens}}=\mathrm{Tr}\left[ \rho _{0}\sum_{n=0}^{\infty }%
\frac{(-i)^{n}}{n!}\int_{\text{c}}dt_{1}\cdots \int_{\text{c}}dt_{n}\mathrm{%
\ }\mathscr{T}_{\mathrm{c}}H(t_{1})\cdots {H{}}(t_{n})\right] ,
\label{Series_ensemble}
\end{equation}%
where each term is the sum of ensemble-averaged
$\mathscr{T}_{\mathrm{c}}$-products of spin operators. Consider the ensemble average
$\left\langle F_{n}\right\rangle \equiv \mathrm{Tr}\left[ \rho _{0}F_{n}\right] $ of an
arbitrary product $F_{n}\equiv \mathscr{T}_{\mathrm{c}}J_{m_{1}}^{\alpha_{1}}(t_{1})
\cdots J_{m_{n}}^{\alpha _{n}}(t_{n})$, where the subscripts
$m_{q}\in \left\{ 1,2,\cdots ,N\right\} $ label bath spins and the
superscripts $\alpha _{q}\in \left\{ z,+,-\right\} $ label the spin
operators. In order for $\left\langle F_{n}\right\rangle $ not to vanish,
$J^{+}$ and $J^{-}$ operators must make pairs. An $F_{n}$ consisting of
$J^{z}$ operators only is called a fully contracted product, whose
ensemble average is trivially evaluated. So next we consider the case with
at least one $J^{+}$ operator,
\begin{equation*}
F_{n}=\mathscr{T}_{\mathrm{c}}J_{m_{1}}^{\alpha _{1}}(t_{1})\cdots
J_{m_{k-1}}^{\alpha _{k-1}}(t_{k-1})J_{m_{k}}^{+}(t_{k})J_{m_{k+1}}^{\alpha
_{k+1}}\cdots J_{m_{n}}^{\alpha _{n}}(t_{n}).
\end{equation*}%
For the moment, we assume that the product in $F_{n}$ is already
time-ordered with $t_{1}>t_{2}>\cdots >t_{n}$ and use the following
procedure to reduce $\left\langle F_{n}\right\rangle $ to the sum of
ensemble-averaged fully contracted products. First we move the spin raising
operator $J_{m_{k}}^{+}(t_{k})$ to the right, generating commutators
between $J_{m_{k}}^{+}(t_{k})$ and the operators on its right,
\begin{align*}
\left\langle F_{n}\right\rangle & =\mathrm{Tr}\left[ \rho
_{0}\sum_{p=k+1}^{n}J_{m_{1}}^{\alpha _{1}}(t_{1})\cdots \left[
J_{m_{k}}^{+}(t_{k}),J_{m_{p}}^{\alpha _{p}}(t_{p})\right] \cdots
J_{m_{n}}^{\alpha _{n}}(t_{n})\right] \\
& +\mathrm{Tr}\left[ \rho _{0}J_{m_{1}}^{\alpha _{1}}(t_{1})\cdots
J_{m_{k-1}}^{\alpha _{k-1}}(t_{k-1})J_{m_{k+1}}^{\alpha
_{k+1}}(t_{k+1})\cdots J_{m_{n}}^{\alpha _{n}}(t_{n})J_{m_{k}}^{+}(t_{k})%
\right] .
\end{align*}%
Then in the second term, we use the cyclic invariance of the trace to move $%
J_{m_{k}}^{+}(t_{k})$ to the left (before $\rho _{0}$), and use $%
J_{m}^{+}(t)\rho _{0}=e^{\beta \omega _{m}}\rho _{0}J_{m}^{+}(t)$ to move $%
J_{m_{k}}^{+}(t_{k})$ across $\rho _{0}$ back to its original position to
get $e^{\beta \omega _{m_{k}}}\left\langle F_{n}\right\rangle $. During this
process, we get additional commutators $e^{\beta \omega _{m_{k}}}\mathrm{Tr}%
\left[ \rho _{0}\sum_{p=1}^{k-1}J_{m_{1}}^{\alpha _{1}}(t_{1})\cdots \left[
J_{m_{k}}^{+}(t_{k}),J_{m_{p}}^{\alpha _{p}}(t_{p})\right] \cdots
J_{m_{n}}^{\alpha _{n}}(t_{n})\right] .$ Collecting all terms, we obtain
\begin{equation}
\left\langle F_{n}\right\rangle =\mathrm{Tr}\left[ \rho _{0}\sum_{p(\neq
k)}J_{m_{1}}^{\alpha _{1}}(t_{1})\cdots \left[ J_{m_{p}}^{\alpha _{p}}(t_{p})%
\right] ^{\bullet }\left[ J_{m_{k}}^{+}(t_{k})\right] ^{\bullet }\cdots
J_{m_{n}}^{\alpha _{n}}(t_{n})\right] ,  \label{Lemma_ensembleA}
\end{equation}%
where the contraction between a spin raising operator $J_{m}^{+}(t)$ and an
arbitrary spin operator $J_{n}^{\alpha }(t_{\alpha })$ is defined as
\begin{align*}
\left[ J_{n}^{\alpha }(t_{\alpha })\right] ^{\bullet }\left[ J_{m}^{+}(t)%
\right] ^{\bullet }& \equiv \left[ J_{m}^{+}(t)\right] ^{\bullet }\left[
J_{n}^{\alpha }(t_{\alpha })\right] ^{\bullet } \\
& \equiv \delta _{m,n}G_{m}(t_{\alpha }-t)\left[ J_{n}^{\alpha },J_{m}^{+}%
\right] (t_{\alpha }),
\end{align*}%
with $G_{m}(t)\equiv \theta (t)\left[ 1+f(\omega _{m})\right] +\theta
(-t)f(\omega _{m})$ being a Green's function and $f(\omega )=1/(e^{\beta
\omega }-1)$. Note that the contraction $[J_{n}^{\alpha }(t_{\alpha
})]^{\bullet }[J_{m}^{+}(t)]^{\bullet }\sim \left[ J_{n}^{\alpha },J_{m}^{+}%
\right] (t_{\alpha })$ is still an operator associated with a contour time $%
t_{\alpha }$ and should be used in subsequent contractions. With $\left[
J^{z},J^{+}\right] =J^{+}$ and $\left[ J^{-},J^{+}\right] =-2J^{z},$ the
contraction between $J_{m}^{+}(t)$ and $J_{m}^{z}(t_{\alpha })$ [or
$J_{m}^{-}(t_{\alpha })$] eliminates $J_{m}^{+}(t)$ and converts
$J_{m}^{z}(t_{\alpha })$ [or $J_{m}^{-}(t_{\alpha })$] to
$J_{m}^{+}(t_{\alpha })$ [or $-2J_{m}^{z}(t_{\alpha })$], reducing the number
of spin operators by one. The resulting product in Eq.~(\ref{Lemma_ensembleA})
is still in a time-ordered sequence, so the time-ordering operator can be
recovered so that
\begin{align}
& \mathrm{Tr}\left[ \rho _{0}\mathscr{T}_{\mathrm{c}}J_{m_{1}}^{\alpha
_{1}}(t_{1})\cdots J_{m_{k-1}}^{\alpha
_{k-1}}(t_{k-1})J_{m_{k}}^{+}(t_{k})J_{m_{k+1}}^{\alpha
_{k+1}}(t_{k+1})\cdots J_{m_{n}}^{\alpha _{n}}(t_{n})\right]  \notag \\
& =\mathrm{Tr}\left[ \rho _{0}\sum_{p(\neq k)}\mathscr{T}_{\mathrm{c}%
}J_{m_{1}}^{\alpha _{1}}(t_{1})\cdots \left[ J_{m_{k}}^{+}(t_{k})\right]
^{\bullet }\left[ J_{m_{p}}^{\alpha _{p}}(t_{p})\right] ^{\bullet }\cdots
J_{m_{n}}^{\alpha _{n}}(t_{n})\right] .  \label{Lemma_ensemble}
\end{align}%
In the above equation the time ordering $t_{1}>t_{2}>\cdots >t_{n}$ is no
longer assumed, since all the spin
operators commute in the $\mathscr{T}_{\mathrm{c}}$-product.

The contraction procedure in Eq.~(\ref{Lemma_ensemble}) can be
repeated whenever there is still a spin raising operator left. Thus
Wick's theorem
follows: Under the average over a noninteracting thermal ensemble, a $%
\mathscr{T}_{\mathrm{c}}$-product of spin operators can be replaced by the
sum of all possible fully contracted products which contains only $J^{z}$
operators.


\begin{figure}[ptb]
\includegraphics[width=\columnwidth]{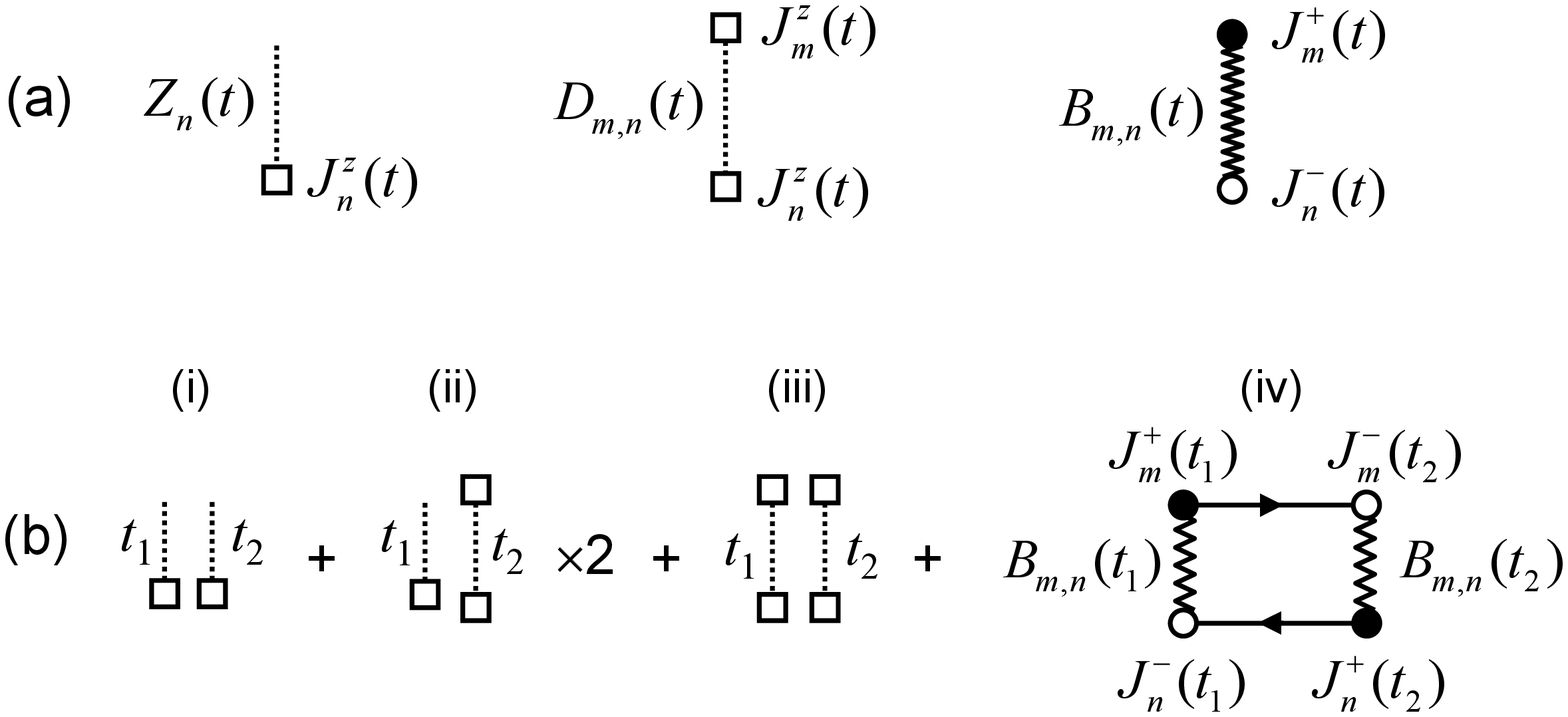}
\caption{(a) Diagrammatic representation of the single-spin term $%
Z_{n}(t)J_{n}^{z}(t),$ the diagonal interaction $%
D_{m,n}(t)J_{m}^{z}(t)J_{n}^{z}(t),$ and the off-diagonal interaction $%
B_{m,n}(t)J_{m}^{+}(t)J_{n}^{-}(t).$ (b) Second-order fully contracted
diagrams.}
\label{G_D12}
\end{figure}

According to Wick's theorem, each term in Eq.~(\ref{Series_ensemble})
generates a series of fully contracted products. The fully contracted
products can be visualized by Feynman diagrams with the following definition
of constituent elements and construction rules.
\begin{enumerate}
\item A spin operator $J^{+}$, $J^{-}$, $J^{z}$ is represented by a vertex
as a filled circle, an empty circle, or an empty square, respectively;

\item Each diagonal (off-diagonal) interaction term containing $n$ spin
operators in the Hamiltonian is represented by a dashed (wavy) interaction
line connecting $n$ vertices;

\item Each contraction $\left[J_{m}^{+}(t)\right]^{\bullet}\left[%
J_{m}^{\alpha}(t_{\alpha})\right]^{\bullet}$ is represented by a
solid arrow starting from the vertex $J_{m}^{+}(t)$ and ending at the vertex $%
J_{m}^{\alpha}(t_{\alpha})$. At the end of the propagating arrow the
commutator $\left[ J_{m}^{\alpha},J_{m}^{+}\right]$ is to be taken;

\item Each $J^{+}(t)$ vertex denoted by a filled circle is connected with
one outgoing propagating arrow, each $J^{z}(t)$ vertex denoted by an empty
square is either free-standing or connected to one incoming arrow
[converting $J^{z}(t)$ to $J^{+}(t)$] and one outgoing arrow [from the
resulting operator $J^{+}(t)$], and each $J^{-}(t)$ vertex denoted by an
empty circle is connected to one incoming arrow [converting $J^{-}(t)$ to $%
-2J^{z}(t)$] or two incoming arrows [the first arrow converting
$J^{-}(t)$ to $-2J^{z}(t)$, and the second arrow converting
$-2J^{z}(t)$ to $-2J^{+}(t)$] and one outgoing arrow [from the
resulting $J^{+}(t)$].
\end{enumerate}
Note that each fully contracted product (and hence each diagram) is an
operator consisting of $J^{z}$ spin operators only.

Taking the Hamiltonian in Eq.~(\ref{HtExample}) for example, the diagonal
single-spin term $Z _{n}(t)J_{n}^{z}(t),$ the diagonal interaction $%
D_{m,n}(t)J_{m}^{z}(t)J_{n}^{z}(t),$ and the off-diagonal interaction
$B_{m,n}(t)J_{m}^{+}(t)J_{n}^{-}(t)$ are visualized in Fig.~\ref{G_D12}(a).
The first-order expansion $(-i)\int_{\text{c}}dt\ {H{}}(t)$ in
Eq.~(\ref{Series_ensemble}) gives two fully contracted products [the first two
diagrams in Fig.~\ref{G_D12}(a)]. The second-order expansion of
Eq.~(\ref{Series_ensemble}) gives four fully contracted products shown in
Fig.~\ref{G_D12}(b). The first three diagrams
\begin{align*}
(\mathrm{i})& =\frac{(-i)^{2}}{2!}Z
_{m}(t_{1})Z_{n}(t_{2})J_{m}^{z}(t_{1})J_{n}^{z}(t_{2}), \\
(\mathrm{ii})& =2\times \frac{(-i)^{2}}{2!}%
Z_{m}(t_{1})D_{n,l}(t_{2})J_{m}^{z}(t_{1})J_{n}^{z}(t_{2})J_{l}^{z}(t_{2}),
\\
(\mathrm{iii})& =\frac{(-i)^{2}}{2!}%
D_{m,n}(t_{1})D_{p,q}(t_{2})J_{m}^{z}(t_{1})J_{n}^{z}(t_{1})J_{p}^{z}(t_{2})J_{q}^{z}(t_{2}),
\end{align*}%
come from diagonal terms and involve no contractions. Here, as a convention,
we have suppressed the sum over spin indices and the contour time integrals.
The last diagram
\begin{align}
(\mathrm{iv})=& \frac{(-i)^{2}}{2!}B_{m,n}(t_{1})B_{m,n}(t_{2})  \notag \\
& \left[ J_{m}^{+}(t_{1})\right] ^{\diamond }\left[ J_{n}^{-}(t_{1})\right]
^{\bullet }\left[ J_{n}^{+}(t_{2})\right] ^{\bullet }\left[ J_{m}^{-}(t_{2})%
\right] ^{\diamond }  \label{SecondOrderExp}
\end{align}
comes from two off-diagonal interaction terms and involves two contractions
$\left[ J_{m}^{+}(t_{1})\right] ^{\diamond }\left[
J_{m}^{-}(t_{2})\right] ^{\diamond }$ and $\left[
J_{n}^{-}(t_{1})\right] ^{\bullet }\left[ J_{n}^{+}(t_{2})\right]
^{\bullet }$, as indicated by the two solid arrows in
Fig.~\ref{G_D12}~(b)~(iv).

\begin{figure}[tbp]
\includegraphics[width=\columnwidth]{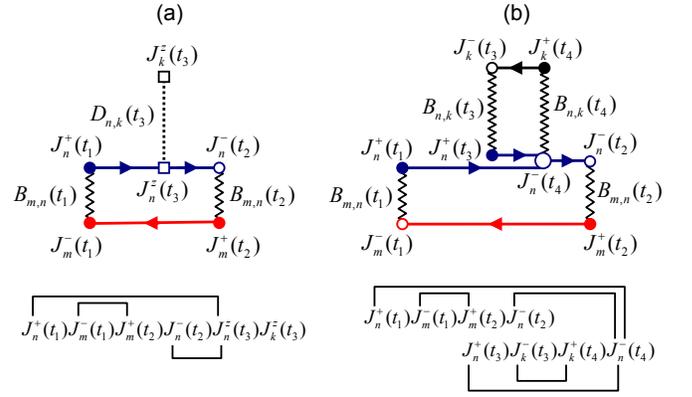}
\caption{(Color online) (a) A third-order connected diagram consisting of
one diagonal and two off-diagonal interaction terms. (b) A 4th-order
connected diagram consisting of four off-diagonal interactions. The
contraction processes contained in each diagram are given below.}
\label{G_ContractionExample}
\end{figure}

Similarly, higher-order diagrams can be constructed by using the above
Feynman rules. Figure~\ref{G_ContractionExample} gives two examples. The
3rd-order diagram in Fig.~\ref{G_ContractionExample}(a) consists of two
off-diagonal interactions $B_{m,n}(t_{1})J_{n}^{+}(t_{1})J_{m}^{-}(t_{1})$
and $B_{m,n}(t_{2})J_{m}^{+}(t_{2})J_{n}^{-}(t_{2})$ and one diagonal
interaction $D_{n,k}(t_{3})J_{n}^{z}(t_{3})J_{k}^{z}(t_{3}).$ It contains
three contractions, one on spin $\mathbf{J}_{m}$ and two on spin $\mathbf{J}%
_{n}$, corresponding to the three solid arrows. The 4th-order one in Fig.~\ref%
{G_ContractionExample}(b) consists of four off-diagonal interactions
and contains five contractions,
three on spin $\mathbf{J}_{n}$ and one on each of the other two spins.

To illustrate the evaluation of the diagrams, we consider again the secular
pair-interaction Hamiltonian in Eq.~(\ref{HtExample}) as the example. The
rules for constructing the analytical formula for a diagram are:
\begin{enumerate}
\item A contour-time dependent constant is associated with each interaction
line, namely, $Z_{n}(t)$ for an open-ended dashed line representing the spin
splitting, $D_{m,n}(t)$ for a dashed line connecting two diagonal spin
operators representing the diagonal interaction, and $B_{m,n}(t)$ for a wavy
line representing the off-diagonal interaction;

\item Each solid arrow from $J_{m}^{+}(t)$ to $J_{m}^{\alpha }(t_{\alpha })$
gives the Green's function $G_{m}(t_{\alpha }-t)$, each freestanding $%
J_{n}^{z}$ vertex gives $J_{n}^{z}$, each $J_{n}^{-}$ vertex
connected to one incoming arrow gives $(-2)J_{n}^{z}$, and each
$J_{n}^{-}$ vertex connected with two incoming arrows and one
outgoing arrow gives $(-2)$;

\item A global factor $(-i)^{k}/k!$ is associated with a diagram containing $%
k$ interaction lines;

\item The spin indices are summed over and the contour times are integrated
over.
\end{enumerate}
For example, the last diagram in Fig.~\ref{G_D12}(b) gives
\begin{equation*}
\frac{(-i)^{2}}{2!}B_{m,n}\left( t_{1}\right) B_{m,n}\left( t_{2}\right)
G_{m}\left( t_{2}-t_{1}\right) G_{n}\left( t_{1}-t_{2}\right) \left(
-2J_{m}^{z}\right) \left( -2J_{n}^{z}\right) ,
\end{equation*}%
which can also be evaluated directly from Eq.~(\ref{SecondOrderExp}) by
carrying out the two contractions.

\begin{figure}[tbp]
\includegraphics[width=\columnwidth]{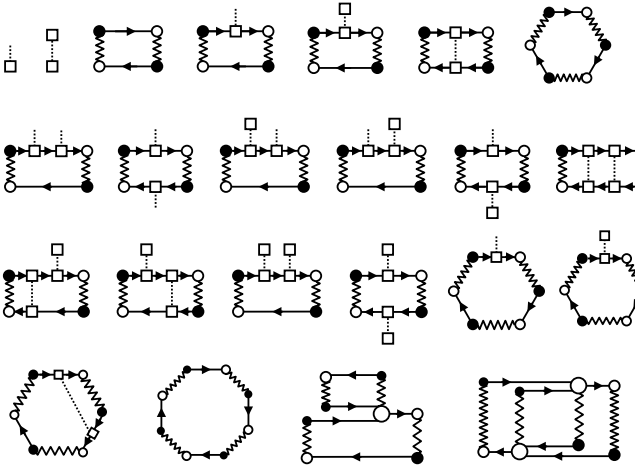}
\caption{Topologically inequivalent connected diagrams up to the 4th order
for the Hamiltonian in Eq.~(\protect\ref{HPN}).}
\label{G_D123}
\end{figure}

Summation of all the fully contracted diagrams leads to the LCE of the
ensemble-averaged evolution
\begin{equation}
\text{T}\mathrm{r}\left( \rho _{0}\mathscr{T}_{\mathrm{c}}e^{-i\int_{\text{c}%
}{H{}}(t)dt}\right) =\text{T}\mathrm{r}\left( \rho _{0}e^{{\hat{\pi}}%
}\right) ,  \label{LCE_ens}
\end{equation}%
where ${\hat{\pi}}$ is the sum of all the \emph{connected} diagrams, such as
the first two diagrams in Fig.~\ref{G_D12}(a) and the last diagram in Fig.~%
\ref{G_D12}(b), but does not include the disconnected ones such as the first
three diagrams in Fig.~\ref{G_D12}(b). As an example, for the Hamiltonian in
Eq.~(\ref{HtExample}), all the topologically inequivalent connected diagrams
up to the 4th order are shown in Fig.~\ref{G_D123}. We see that the number
of diagrams increases significantly with increasing perturbation order.

\subsection{Single-sample LCE for spin-1/2 baths}

\label{append_LCEspin12}

For a single noninteracting bath state $|{\mathcal{J}}\rangle $ (in contrast
to the thermal ensemble), the LCE for spin-1/2 baths has been outlined in
Ref.~[\onlinecite{Saikin2007}]. Here we reproduce the main results using
the ensemble LCE depicted in the previous subsection.

The key is that for a spin-1/2 bath, any noninteracting state $\left\vert
\mathcal{J}\right\rangle =\mathbf{\otimes }_{n}\left\vert j_{n}\right\rangle
$ can be taken as the ground state of a corresponding noninteracting
Hamiltonian,
\begin{equation}
H_{\mathcal{J}}=\sum_{n}\omega _{n}J_{n}^{z},  \label{H_J}
\end{equation}%
where $\omega _{n}<0$ (or $>0$) for $\left\vert j_{n}\right\rangle
=\left\vert \uparrow \right\rangle $ (or $\left\vert \downarrow
\right\rangle $). So the noninteracting single-sample average $\left\langle
\mathcal{J}\left\vert O\right\vert \mathcal{J}\right\rangle $ becomes the
zero-temperature limit ($\beta \rightarrow +\infty $) of the corresponding
noninteracting ensemble average $\mathrm{Tr}\left( \rho _{\mathcal{J}%
}O\right) $ with the density matrix $\rho _{\mathcal{J}}\equiv e^{-\beta H_{%
\mathcal{J}}}/\mathrm{Tr}\left( e^{-\beta H_{\mathcal{J}}}\right) $. In
particular, the single-sample expectation value of the bath propagator
\begin{equation*}
{\mathcal{L}}=\left\langle \mathcal{J}\left\vert \mathscr{T}_{\mathrm{c}%
}e^{-i\int_{\text{c}}H(t)dt}\right\vert \mathcal{J}\right\rangle
=\lim_{\beta \rightarrow +\infty }\mathrm{Tr}\left[ \rho _{\mathcal{J}}%
\mathscr{T}_{\mathrm{c}}e^{-i\int_{\text{c}}H(t)dt}\right] .
\end{equation*}%
Thus the single-sample LCE is obtained by simply setting the Green's
function $G_{n}(t)=\theta (t)\delta _{j_{n},\downarrow }-\theta (-t)\delta
_{j_{n},\uparrow }$ and replacing the ensemble average $\mathrm{Tr(}\rho
_{0}\cdots )$ with $\left\langle \mathcal{J}\left\vert \cdots \right\vert
\mathcal{J}\right\rangle $, i.e.,
\begin{equation}
\mathcal{L}=\left\langle \mathcal{J}\left\vert \mathscr{T}_{\mathrm{c}%
}e^{-i\int_{\text{c}}{H{}}(t)dt}\right\vert \mathcal{J}\right\rangle
=\left\langle \mathcal{J}\left\vert e^{{\hat{\pi}}}\right\vert \mathcal{J}%
\right\rangle .  \label{LCE_single}
\end{equation}%
Note that the connected diagrams in $\hat{\pi}$ contain only $J^{z}$
operators, which commute with each other. Thus the single-sample average can
be performed for each diagram to convert it into a $c$-number, so that
\begin{equation*}
\mathcal{L}=\exp \left( {\left\langle \mathcal{J}\left\vert \hat{\pi}%
\right\vert \mathcal{J}\right\rangle }\right) \equiv \exp \left( {\pi }%
\right) .
\end{equation*}

\subsection{Single-sample LCE for higher spins}

\label{append_LCEgeneral}
\begin{figure}[ptb]
\includegraphics[width=\columnwidth]{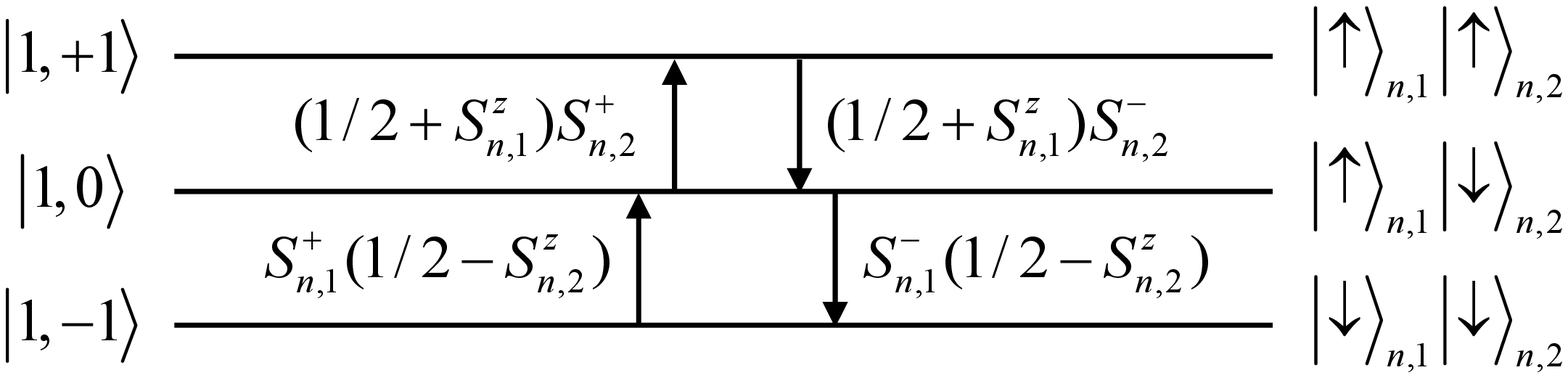}
\caption{Mapping of spin-1 operators $J^{\pm}$ to pseudo-spin-1/2
operators. The so-defined mapping restricts the state evolution of
the two pseudo-spin-1/2's within the physical Hilbert space of the
spin-1.} \label{G_Spin1Mapping}
\end{figure}

\begin{figure}[ptb]
\includegraphics[width=\columnwidth]{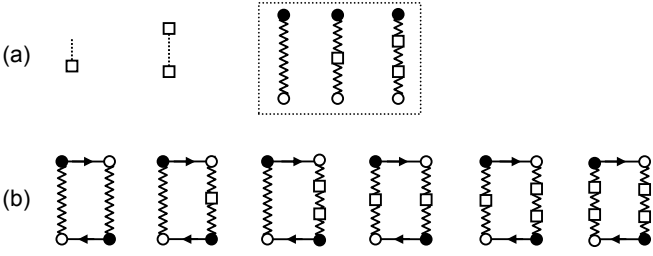}
\caption{ Diagram representation of (a) $Z_{n}J_{n}^{z}$, (b)$%
D_{m,n}J_{m}^{z}J_{n}^{z}$, and (c) $B_{m,n}J_{m}^{+}J_{n}^{-}$. (d)
Topologically inequivalent second-order connected diagrams for a spin-1
bath. }
\label{G_Spin1LD12}
\end{figure}

For a higher-spin bath, a noninteracting single sample state $|{\mathcal{J}}%
\rangle $ in general is not the eigen state of a noninteracting Hamiltonian
as in Eq.~(\ref{H_J}). Thus the single-sample LCE for higher-spin baths may
not be derived from the ensemble LCE directly. Here we provide a solution by
mapping a higher spin to a composite of pseudo-spin-1/2's. The mapping
is such that the physical states form an invariant subspace in the much
larger pseudo-spin Hilbert space.

Without loss of generality, we consider a spin-1 $\mathbf{J}_{n}$.
The mapping from the spin-1 states to the states of two pseudo-spin-1/2's ${\mathbf{S}}%
_{n,\lambda }$ ($\lambda =1,2$) is
\begin{align*}
\left\vert 1,-1\right\rangle _{n}& \rightarrow \left\vert \downarrow
\right\rangle _{n,1}\left\vert \downarrow \right\rangle _{n,2}, \\
\left\vert 1,0\right\rangle _{n}& \rightarrow \left\vert \uparrow
\right\rangle _{n,1}\left\vert \downarrow \right\rangle _{n,2}, \\
\left\vert 1,+1\right\rangle _{n}& \rightarrow \left\vert \uparrow
\right\rangle _{n,1}\left\vert \uparrow \right\rangle _{n,2},
\end{align*}%
as schematically shown in Fig.~\ref{G_Spin1Mapping}. The two
pseudo-spin-1/2's have more basis states than the spin-1. To restrict the
evolution of the spin state within the physical Hilbert space, we map the
spin-1 operators to pseudo-spin-1/2 operators as (see Fig.~\ref%
{G_Spin1Mapping})
\begin{align*}
J_{n}^{+}& \rightarrow \sqrt{2}S_{n,1}^{+}\left( 1/2-S_{n,2}^{z}\right) +%
\sqrt{2}\left( 1/2+S_{n,1}^{z}\right) S_{n,2}^{+}, \\
J_{n}^{-}& \rightarrow \sqrt{2}S_{n,1}^{-}\left( 1/2-S_{n,2}^{z}\right) +%
\sqrt{2}\left( 1/2+S_{n,1}^{z}\right) S_{n,2}^{-}, \\
J_{n}^{z}& \rightarrow S_{n,1}^{z}+S_{n,2}^{z},
\end{align*}%
in which the flip-flop of one pseudo-spin-1/2 is conditioned on the state of
the other pseudo-spin-1/2. Thus each spin-1 operator is mapped to an
interaction term of the two pseudo-spin-1/2's. An interaction term between
two spin-1's would contain up to four pseudo-spin-1/2 operators, making the
Hamiltonian rather complicated. Nonetheless, the LCE can be readily applied
to the spin-1 bath mapped to a pseudo-spin-1/2 one. For a spin-1 Hamiltonian
as in Eq.~(\ref{HtExample}), the different terms $%
Z_{n}(t)J_{n}^{z}(t),D_{m,n}(t)J_{m}^{z}(t)J_{n}^{z}(t),$ and $%
B_{m,n}(t)J_{m}^{+}(t)J_{n}^{-}(t)$ converted to pseudo-spin-1/2 operators,
are represented in turn by the three diagrams in Fig.~\ref{G_Spin1LD12}(a).
There each vertex (denoting a pseudo-spin-1/2 operator according to the same
Feynmann rule depicted in Appendix~\ref{append_LCEGaussian}) is associated
with a physical spin index $n\in \{1,2,\cdots ,N\}$ and a pseudo-spin index $%
\lambda \in \{1,2\}$. The mapping leads to three different types of
off-diagonal interactions (including simultaneous interaction involving up
to four pseudo-spin-1/2's). As a result, the number of connected diagrams
increases dramatically, e.g., we have six topologically inequivalent
second-order connected diagrams for spin-1 baths, as shown in Fig.~\ref%
{G_Spin1LD12}(b), while for a spin-1/2 bath we have only one such
diagram (the third diagram in Fig.~\ref{G_D123}).


\end{document}